\documentclass[%
 reprint,
superscriptaddress,
showpacs,
 amsmath,amssymb,
 aps,
 pra,
 floats,
]{revtex4-1}

\usepackage[latin1]{inputenc}
\usepackage{graphicx}
\usepackage{dcolumn}
\usepackage{bm}
\usepackage{units}
\usepackage{color}

\newcommand{\ket}[1]{ \left|#1\right\rangle}
\begin{document}

\title{Storage of fiber-guided light in a nanofiber-trapped ensemble of cold atoms}%

\author{C. Sayrin}
\author{C. Clausen}
\author{B. Albrecht}
\author{P. Schneeweiss}
\email{Schneeweiss@ati.ac.at}
\author{A. Rauschenbeutel}
\email{Arno.Rauschenbeutel@ati.ac.at}
\affiliation{%
Vienna Center for Quantum Science and Technology, Atominstitut, TU Wien, Stadionallee 2, 1020 Vienna, Austria
}%

\date{\today}


\begin{abstract}
Tapered optical fibers with a nanofiber waist are versatile tools for interfacing light and matter. In this context, laser-cooled atoms trapped in the evanescent field surrounding the optical nanofiber are of particular interest: They exhibit both long ground-state coherence times and efficient coupling to fiber-guided fields. Here, we demonstrate electromagnetically induced transparency, slow light, and the storage of fiber-guided optical pulses in an ensemble of cold atoms trapped in a nanofiber-based optical lattice. We measure a slow-down of light pulses to group velocities of 50~m/s. Moreover, we store optical pulses at the single photon level and retrieve them on demand in the fiber after 2~$\mu$s with an overall efficiency of ($3.0\pm0.4$)~\%. Our results show that nanofiber-based interfaces for cold atoms have great potential for the realization of building blocks for future optical quantum information networks.
\end{abstract}

\maketitle


Storing classical light pulses in optical memories is an important capability for the realization of all-optical signal processing schemes. Similarly, quantum information processing and communication require quantum memories in which quantum states of light can be faithfully stored~\cite{Lvovsky09,Simon10}. Such memories are crucial elements of future large-scale quantum optical networks~\cite{Bussieres13a}. They are key to quantum repeaters~\cite{Briegel98} which are indispensable when it comes to the exchange of quantum information over long-distances~\cite{Duan01,Sangouard11,Northup14}. Furthermore, quantum memories can be used to effectively turn a probabilistic single-photon source into an on-demand source~\cite{Nunn13}.

A classical light pulse was stored for about one minute in a rare-earth-doped crystal using dynamical decoupling techniques~\cite{Heinze13}. Similar storage times were achieved with an ensemble of ultracold atoms in a free-space optical lattice~\cite{Dudin13}. For optical network-based applications, efficient and long-lived fiber-integrated optical memories are desirable and currently constitute an active field of research~\cite{Bajcsy09,Sprague14,Saglamyurek15}. Recently, weak coherent light pulses were stored using a hollow-core photonic-crystal fiber filled with thermal cesium vapor~\cite{Sprague14}. In this case, the highest measured memory efficiency was 27~\% and the memory lifetime was about 30~ns.  In another recent work, photons at a wavelength of about 1.5~$\mu$m were stored and retrieved with an efficiency of 1~\% after 5~ns using a cryogenically cooled erbium-doped fiber~\cite{Saglamyurek15}. There, it was also demonstrated that photonic entanglement is preserved during storage and retrieval. The performance of these fiber-integrated quantum memories could be significantly improved if decoherence mechanisms such as the motion of the atoms or the coupling to the solid-state environment were suppressed.

Here, we make use of a nanofiber-based optical interface~\cite{Vetsch10} to store fiber-guided light in an ensemble of trapped neutral atoms. The laser-cooled cesium atoms are confined in a one-dimensional optical lattice that is realized in the evanescent field surrounding an optical nanofiber. The latter forms the waist of a tapered optical fiber (TOF) which enables close-to-unity coupling of light fields that are guided in a standard optical fiber into and out of the nanofiber section. Our experiments rely on electromagnetically-induced transparency (EIT) where both the probe and the control light fields are nanofiber-guided and couple to the atoms via their evanescent fields. We observe a spectrally narrow transparency window with a width of about 30~kHz which reaches more than 60~\% transmission through the TOF with an otherwise optically dense atomic ensemble. We study the probe pulse propagation under EIT conditions and demonstrate slow light with a group velocity of about 50~m/s. Finally, we store fiber-guided light pulses at the single-photon level for $2~\mu$s and retrieve them with a combined efficiency of ($3.0\pm0.4$)~\%.

\begin{figure}[t]
	\centering
	\includegraphics[width=1.0\columnwidth]{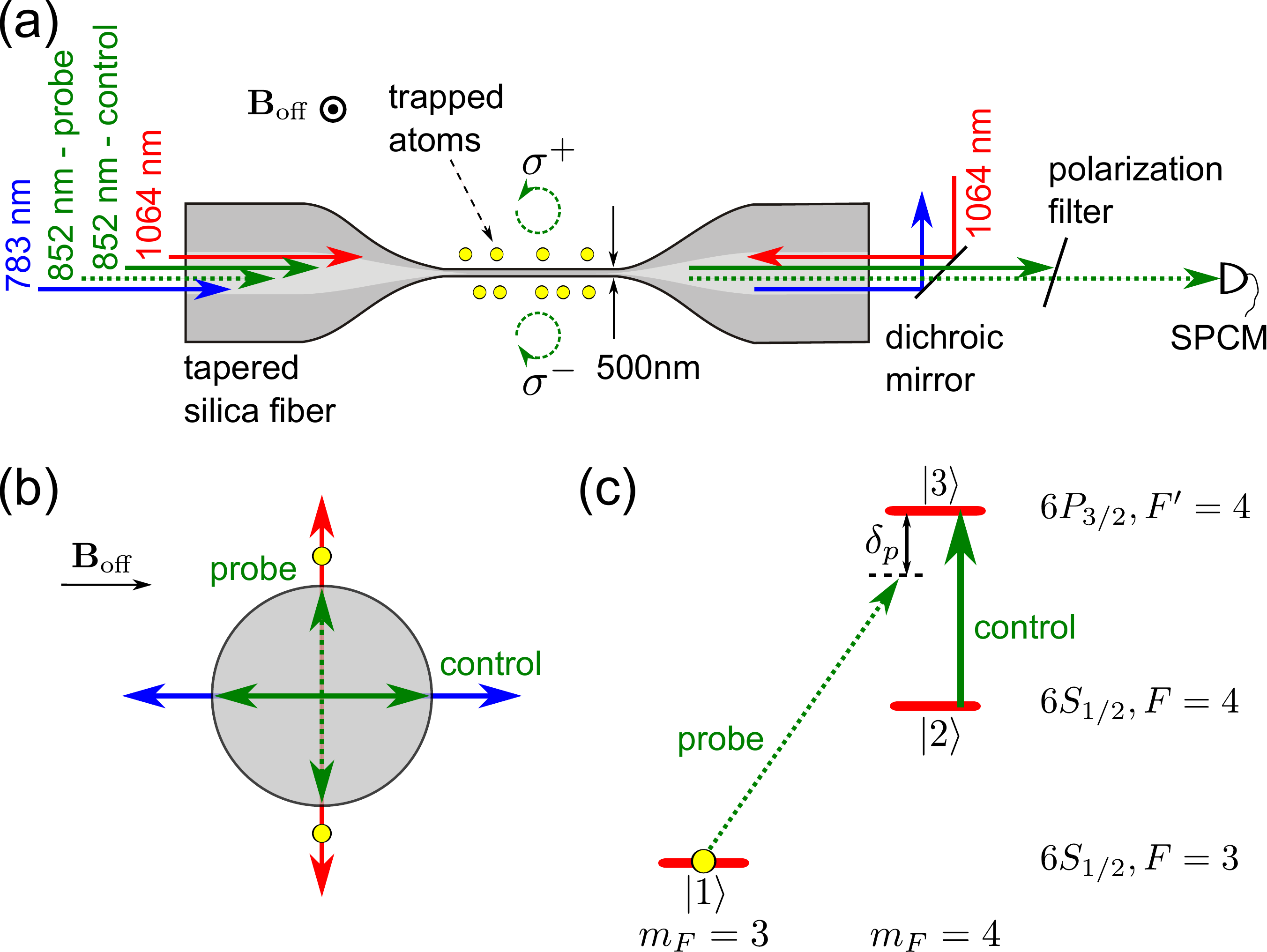}
	\caption{(a) Schematic of the experimental setup including the tapered optical fiber, the trapping, probe, and control laser fields, and the single-photon counting module (SPCM). The polarization of the probe field above and below the nanofiber is indicated by the dashed circular arrows. A polarization filter in front of the SPCM separates the control laser from the probe field. A homogeneous magnetic field $\mathbf{B}_{\rm off}$ is applied along the indicated direction. (b) Cross-sectional view of the nanofiber, illustrating the orientations of the principle axes of quasi-linear polarizations of the nanofiber-guided fields. (c) Relevant Zeeman sublevels of the trapped cesium atoms. The states $\ket{1}$, $\ket{3}$, and $\ket{2}$ form the $\Lambda$-system used in the experiments. The transitions driven by the probe and the control laser fields are also indicated. The quantization axis is chosen along the direction of the magnetic field.}
	\label{fig:setup}
\end{figure}

A schematic of the experimental setup is shown in Fig.~\ref{fig:setup}(a). The silica nanofiber has a radius of $250$~nm, meaning that for all guided light fields used in the experiment the single-mode condition is fulfilled. The nanofiber-based trap is created by sending a blue-detuned running-wave field with a free-space wavelength of $783$~nm and a power of $7.0$~mW and a red-detuned standing-wave field at $1064$-nm wavelength with a power of $0.65$~mW per beam into the nanofiber. The trapping potential consists of two diametric linear arrays of individual trapping sites along the nanofiber, located at a distance of $225$~nm  from the surface. Each site contains at most one atom and offers sub-wavelength confinement in all three spatial dimensions~\cite{Vetsch12}. In the context of light storage, these trap properties are highly advantageous because collisional broadening of atomic transitions is absent and motional dephasing is strongly suppressed.

We realize our experiments using the cesium hyperfine levels shown in Fig.~\ref{fig:setup}(c). The quantization axis is chosen along the applied homogeneous magnetic field $\mathbf{B}_{\rm off}$ indicated in Fig.~\ref{fig:setup}(a). The $\Lambda$-system that is required for EIT is formed with the two Zeeman ground states $\ket{1}=\ket{6S_{1/2}, F=3, m_F=+3}$ and $\ket{2}=\ket{6S_{1/2}, F=4, m_F=+4}$, and the Zeeman excited state $\ket{3}=\ket{6P_{3/2}, F'=4, m_{F'}=+4}$. The probe field which couples the states $\ket{1}$ and $\ket{3}$ as well as the control field which couples the states $\ket{2}$ and $\ket{3}$ are launched into the TOF. They are co-propagating and are both quasi-linearly polarized~\cite{LeKien04b} in the waist of the TOF. Their principle polarization axes are aligned as shown in Fig.~\ref{fig:setup}(b). We take advantage of the particular polarization properties of the nanofiber-guided light fields~\cite{LeKien04b}: At the position of the atoms, the control laser field is $\pi$-polarized. At the same time, the probe light field is almost perfectly $\sigma^+$ ($\sigma^-$)-polarized above (below) the nanofiber in Fig.~\ref{fig:setup}(a). The Zeeman shifts induced by $\mathbf{B}_{\rm off}$ ensure that the probe light field almost exclusively couples to the atoms above the nanofiber~\cite{Mitsch14a}. In the following, the control light field is always resonant with the $\ket{2}\to \ket{3}$ transition. The probe light field of frequency $\omega_{\rm p}$ is phase-locked to the control light field and detuned by $\delta_{\rm p}=\omega_{\rm p}-\omega_{31}$ from the $\ket{1}\to \ket{3}$ transition at frequency~$\omega_{31}$.

\begin{figure}
	\includegraphics[width=1.0\columnwidth]{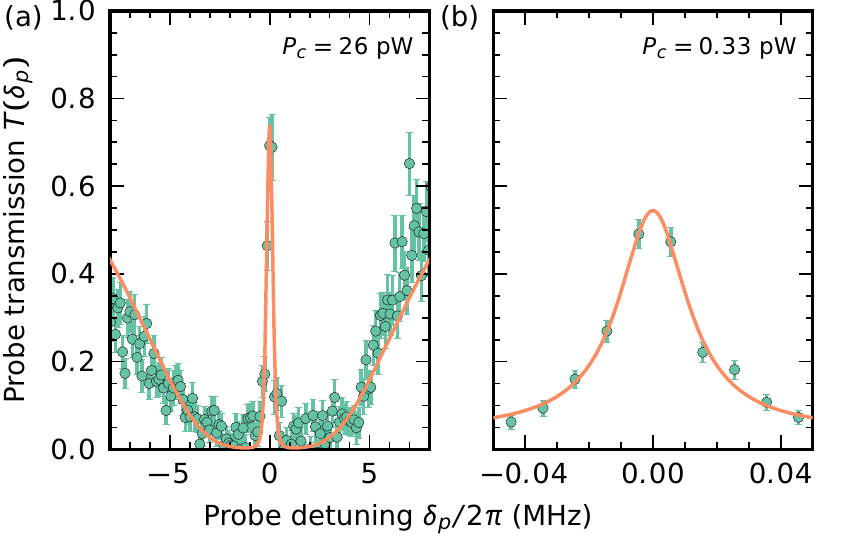}
	\caption{Transmission spectrum of the guided probe field under EIT conditions. (a) A narrow transmission window with a width clearly smaller than the natural linewidth is observed on an optically dense background. The control power is $P_{\rm c}=26$~pW and the probe power is $P_{\rm p}\approx 2.9$~pW. The orange line is a fit to the data based on Eq.~(\ref{eq:T}). The spectrum is averaged over 300 measurements. (b) For a smaller $P_{\rm c}$ of 0.33~pW, we observe a transmission window which is about 10 times narrower. The orange line is the result of a Lorentzian fit. Here, $P_{\rm p}=1.7$~pW. Each data point is an average over 60 measurements. The error bars in (a) and (b) correspond to one standard deviation based on counting statistics.
}
	\label{fig:EIT}
\end{figure}

We measure the probe transmission through the nanofiber under EIT conditions. After loading atoms into the nanofiber-based trap in their $F=3$ hyperfine ground state, we slowly increase $B_{\rm off}$ from 0~G to $26$~G. Through adiabatic magnetization, this process prepares most of the atoms in the state $\ket{1}$~\cite{Sayrin15}. We then switch on the probe and the control fields, and scan the detuning $\delta_{\rm p}$ over 60~MHz within $500~\mu$s. The transmission spectrum $T(\delta_{\rm p})$ is determined by measuring the transmitted power with and without atoms using a single photon counting module (SPCM) with a bin size of 1~$\mu$s. Figure~\ref{fig:EIT}(a) shows the resulting transmission spectrum for a control power of $P_{\rm c}=26$~pW. The probe power is $P_{\rm p}=2.9$~pW. We observe a narrow EIT transmission window with a full width at half maximum (FWHM) of about 300~kHz and a maximum transmission of about 70~\% at resonance, $\delta_{\rm p}=0$. We model the probe transmission as
\begin{equation}
T(\delta_{\rm p}) = {\left|h(\delta_{\rm p})\right|}^2~,\label{eq:T}
\end{equation}
with the transfer function~\cite{Fleischhauer05}
\begin{equation}
h(\delta_{\rm p}) =
		\exp \left( i\,\eta\,\tilde\chi(\delta_{\rm p})/2 \right)~, \label{eq:t}\end{equation}
where
\begin{align}
\tilde\chi (\delta_{\rm p}) &=
		\frac{2\frac{\delta_{\rm p}}{\gamma_{21}} \left(\frac{4\delta_{\rm p}^2-|\Omega_{\rm c}|^2}{\gamma_{21}\gamma_{31}}\right)}{\left|\frac{|\Omega_{\rm c}|^2}{\gamma_{21}\gamma_{31}}+\left(1-2i\frac{\delta_{\rm p}}{\gamma_{21}}\right)\left(1-2i\frac{\delta_{\rm p}}{\gamma_{31}}\right)\right|^2}
\nonumber \\
&+i \frac{4\left(\frac{\delta_{\rm p}}{\gamma_{21}}\right)^2+1+\frac{|\Omega_{\rm c}|^2}{\gamma_{21}\gamma_{31}}}{\left|\frac{|\Omega_{\rm c}|^2}{\gamma_{21}\gamma_{31}}+\left(1-2i\frac{\delta_{\rm p}}{\gamma_{21}}\right)\left(1-2i\frac{\delta_{\rm p}}{\gamma_{31}}\right)\right|^2}~.
\end{align}
Here, $\Omega_{\rm c}$ is the control laser Rabi frequency, $\gamma_{31}$ the excited state decay rate and $\gamma_{21}$ the decay rate of the coherence between the two ground states of the $\Lambda$-system. The quantity $\eta$ is the resonant optical depth in the absence of a control field. The obtained fit function, shown as an orange line in Fig.~\ref{fig:EIT}(a), agrees well with the data: Fixing $\gamma_{31}/(2\pi)=6.4$~MHz based on independent measurements~\cite{Mitsch14a}, we obtain $\gamma_{21}/(2\pi)=(49\pm 18)$~kHz, $\Omega_{\rm c}/(2\pi)=(2.4\pm0.1)$~MHz, and $\eta=5.9\pm0.2$. Given the high absorption per atom of $3.8\,\%$ \cite{LeKien14b}, the measured optical depth corresponds to the contribution of only 160 atoms in the state $\ket{1}$.

The value of $\gamma_{21}$ matches our expectations based on microwave spectroscopy on the ground state manifold~\cite{Mitsch14a} and is promising for light storage. Moreover, it is about one order of magnitude smaller than the measured width of the EIT window. The minimal achievable width of the transmission window is on the order of $\gamma_{21}/\sqrt{\eta}$~\cite{Fleischhauer05}. We confirm this experimentally in Fig.~\ref{fig:EIT}(b), where we set the control power to $P_{\rm c}=0.33$~pW. Here, $T(\delta_{\mathrm{p}})$ is measured separately for each detuning by recording the transmitted power of the probe light field during $10~\mu$s. The power of the probe field, $P_{\rm p}=1.7$~pW, is now comparable to the control field power. After the probe and the control field have been switched on, we thus wait $55~\mu$s until the atomic ensemble is pumped into the dark state~\cite{Arimondo96}. We then measure a transmission window with a Lorentzian FWHM of only 26~kHz and a transmission of about 60~\%.

\begin{figure}[t]
\includegraphics[width=1.0\columnwidth]{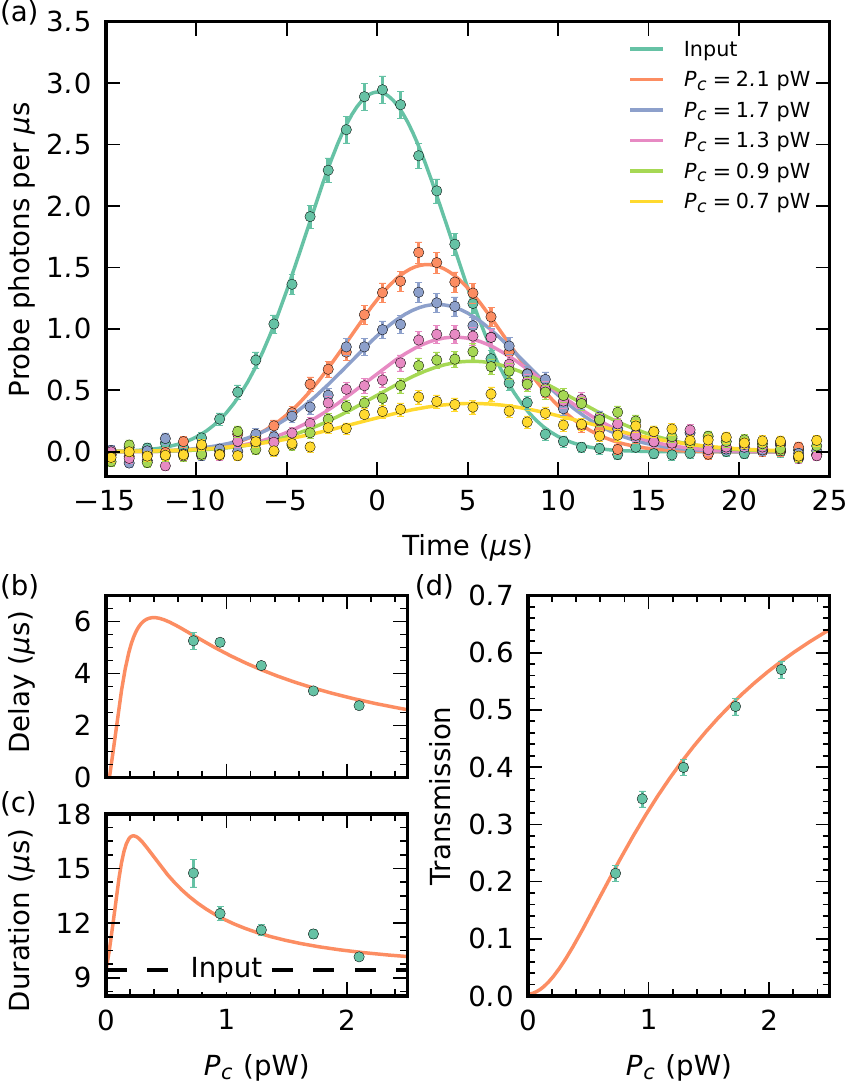}
	\caption{Slow fiber-guided light. (a) Time traces of probe pulses transmitted through the atom--nanofiber system under EIT conditions. A delay of the pulses with respect to a reference pulse (dark green) is clearly visible. The solid lines are Gaussian fits to the data. Each point is the result of the average over 5 pulses per atomic ensemble and over 800 experimental runs. The bin size is 1~$\mu$s. The error bars are as in Fig.~\ref{fig:EIT}. (b) Pulse delay, (c) pulse duration, and (d) pulse transmission as a function of the power of the control light field, $P_{\rm c}$. The orange lines are the results of a global fit of the data sets in (b)--(d), see text. The error bars are the standard errors of the Gaussian fits. The horizontal dashed line in (c) indicates the duration $\tau=9.4~\mu$s of the input pulse.}
	\label{fig:slowlight}
\end{figure}

The observed narrow EIT window implies a steep modulation of the refractive index of the atomic medium around $\delta_{\rm p}=0$. This results in a significant reduction of the group velocity of the probe pulse, i.e., in slow light~\cite{Kasapi95}. We study the propagation of a resonant probe pulse through the TOF under EIT conditions. To this end, we launch a probe pulse with Gaussian temporal profile and a duration of $\tau=9.4~\mu$s (FWHM) into the TOF. The pulse contains 30 photons in the fiber on average, corresponding to a peak power of $0.7$~pW. We record the transmitted probe pulse using an SPCM and compare it to a reference pulse taken in the absence of atoms. This is repeated 5 times with each ensemble of atoms and for 800 experimental runs. The results are shown in Fig.~\ref{fig:slowlight}(a). For example, for a control power of $P_{\rm c}=2.1$~pW, we observe a pulse delay of about 3~$\mu$s.

We repeat this measurement for different values of $P_{\rm c}$. For each power, we infer the delay, duration, and transmission of the probe pulse using a Gaussian fit, see Fig.~\ref{fig:slowlight}(b), (c), and (d), respectively.
We observe that for decreasing $P_{\rm c}$ the delay of the pulse increases. At the same time, the transmitted power is reduced and the pulse duration grows. This fully matches the expected behavior: The spectral width of the EIT window is reduced with decreasing $P_{\rm c}$. The associated steeper modulation of the refractive index leads to smaller group velocities. At the same time, more and more of the pulse spectrum falls outside of the EIT window, leading to both absorption and distortion. We simulate the experiment using the transfer function $h(\delta_{\rm p})$ in Eq.~(\ref{eq:t}) and leaving the optical depth, a common scaling factor for all control field Rabi frequencies, and $\gamma_{21}$ as adjustable parameters. We then fit the outcome of the simulation to the experimental data, see orange lines in Figs.~\ref{fig:slowlight}(b)--(d). The agreement with the data is very good for the following parameters: $\gamma_{21}/(2\pi)=(20\pm 2)$~kHz, consistent with the analysis in Fig.~\ref{fig:EIT}(a); $\Omega_{\rm c}/(2\pi)=(353\pm 26)$~kHz for $P_{\rm c}=1$~pW, close to the prediction of about 290~kHz; and $\eta=6.0\pm 0.7$. Given these parameters, we expect that even longer delays can be observed by lowering $P_{\rm c}$ while simultaneously increasing the pulse duration. We confirmed this by launching pulses of duration $\tau=93\,\mu$s into the medium with $P_{\rm c}=0.33$~pW. In this case, we find a delay of $(22\pm1)~\mu$s at a transmission of $(13.6\pm0.5)\,\%$. Given the length of our atomic sample of about 1~mm, this corresponds to a group velocity of 50~m/s.

\begin{figure}
\includegraphics[width=1.0\columnwidth]{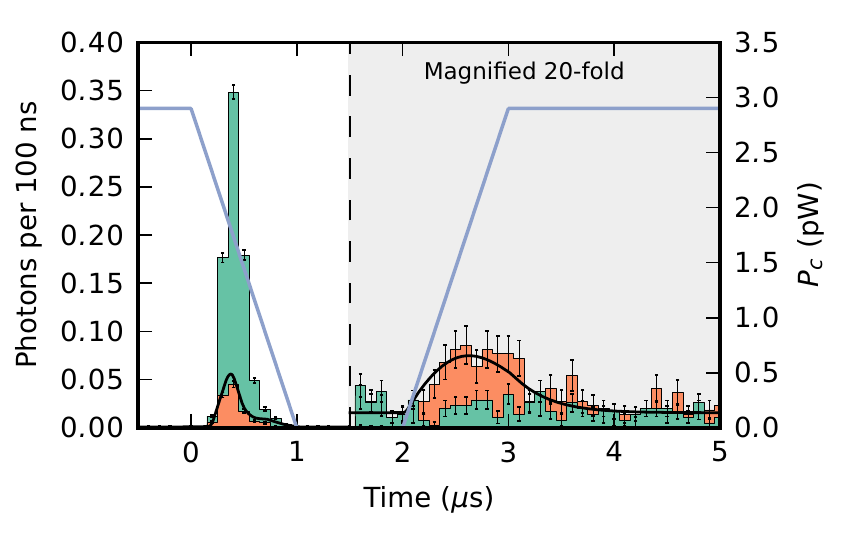}
	\caption{Storage of light in a nanofiber-trapped ensemble of cold atoms. A pulse of duration $\tau=0.2~\mu$s that contains 0.8 photons on average is launched into the TOF and stopped inside the atomic medium. This is achieved by reducing the control laser power $P_{\rm c}(t)$ to zero (blue line). After $1~\mu$s, $P_{\rm c}(t)$ is increased to its initial value, and the pulse is retrieved and recorded by the SPCM. Here, $B_{\rm off}=15$~G, the bin size is 100~ns, and the data is averaged over 55 pulses per atomic ensemble and over 1600 experimental runs (2400 for reference).}
	\label{fig:storage}
\end{figure}

The slow light technique can be readily extended to storage and on-demand retrieval of light pulses. In order to stop the pulse, the control power is reduced to zero while the light propagates through the medium. The retrieval of the light pulse is triggered by switching the control field back on. We launch a pulse of duration $\tau=0.2~\mu$s that contains 0.8 photons on average into the TOF. The control power $P_{\rm c}$ is ramped down linearly with a timing indicated in Fig.~\ref{fig:storage}. After a holding time of $1~\mu$s, we ramp up $P_{\rm c}$ to its initial value. The stored pulse is then released from the medium and detected with the SPCM. This procedure is repeated 55 times with each atomic ensemble and for 1600 experimental runs. The resulting time trace is shown in Fig.~\ref{fig:storage}. The reference pulse is recorded in the same way by repeating the experiment without atoms. We find that the light pulse is retrieved with about $2~\mu$s delay with respect to the reference pulse. The measured combined storage and retrieval efficiency of our memory reaches ($3.0\pm0.4$)~\%.

Under ideal conditions, and in particular in the absence of decoherence, the maximum attainable storage and retrieval efficiency depends only on the optical depth $\eta$~\cite{Gorshkov07b}. For $\eta=2.4$ as in this measurement, this optimal efficiency is $13$~\%. In order to identify the limiting factors for the efficiency for our settings, we simulate the experiment according to Ref.~\cite{Gorshkov07b}. We calculate the full time trace of the pulse which is partially directly transmitted, and partially stored and retrieved, see solid black line in Fig.~\ref{fig:storage}. In addition to the measured value of $\eta$, we take the parameter $\gamma_{21}/(2\pi)=20$~kHz as inferred from the data in Fig.~\ref{fig:slowlight}, and $\gamma_{31}/(2\pi)=6.4$~MHz as before. The simulation outcome accurately reproduces the measured time trace of the probe pulse for $\Omega_{\rm c}(t=0)/(2\pi)=2.24$~MHz. Given the value $P_{\rm c}(t=0)$, however, we calculate $\Omega_{\rm c}/(2\pi)\approx 0.6$~MHz. The discrepancy between these two values is still under investigation and might be caused by a drift of the experiment after the calibration of the power of the control field inside the nanofiber. Given the very good agreement between the measured and the simulated probe time traces, however, we use the simulation to identify the mechanisms that decrease the memory efficiency. We find that this reduction is due to approximately equal contributions from ground-state decoherence, insufficient control power, and a non-optimal control field ramp.

The optical depth of the atomic ensemble could be increased by loading more atoms into the trap via a larger spatial overlap of the initial cold atom cloud and the nanofiber. Concerning decoherence, we have shown that the magnetic field-insensitive $m_F=0$ hyperfine ground states exhibit coherence times on the order of a millisecond in our trap~\cite{Reitz13}. Additional cooling of the trapped atoms, e.g., based on EIT~\cite{Morigi00}, Raman~\cite{Kaufman12,Thompson13}, or microwave sideband-cooling~\cite{Foerster09}, as well as techniques to cancel the trap-induced differential light shifts of the ground states can bring further improvement.

For integrating our memory in an optical fiber telecommunication network, optical frequency conversion from the atomic resonance to the telecom band is required. In this context, efficient conversion from the rubidium D lines to a wavelength of 1.5~$\mu$m in laser-cooled rubidium vapor~\cite{Radnaev10} and in a non-linear crystal~\cite{Albrecht14} has been demonstrated. A transfer of these techniques to a fiber-based platform~\cite{Donvalkar14} offers a promising extension for our nanofiber-based optical memory.

For an optical memory to function also in the quantum regime, low noise operation is essential. The waveguide-geometry in conjunction with the large optical depth per atom in our system allows one to work with a small number of atoms as compared to typical free-space experiments. This reduces read-out noise that would stem from imperfect atomic state preparation. In fact, we stored and retrieved classical light pulses that contained less than one photon on average. This indicates that the noise characteristics of our optical memory should also allow one to efficiently store quantum information and entanglement over millisecond time scales in an all-fiber-based architecture.

We acknowledge financial support by the Austrian Science Fund (FWF, SFB NextLite project No. F~4908-N23 and DK CoQuS project No. W~1210-N16). C.S.~acknowledges support by the European Commission (Marie Curie IEF Grant 328545).

\bibliography{Storage}

\end{document}